**Defect-impurity complex induced long-range ferromagnetism in GaN nanowires**

Assa Aravindh. S and Iman S Roqan[*]

*Physical Sciences and Engineering, King Abdullah University of Science and Technology (KAUST), Thuwal, Saudi Arabia.*

Present work investigates the structural, electronic and magnetic properties of wurtzite (0001) GaN nanowires (NWs) doped with Gd and point defects by employing the GGA+$U$ approximation. We find that Ga vacancies ($V_{Ga}$) exhibit lower formation energy compared to N vacancies ($V_N$). Further stabilization of point defects occurs due to the presence of Gd and ambient ferromagnetism (FM) can be stabilized in the NW by the additional positive charge induced by the $V_{Ga}$. Electronic structure analysis shows that $V_{Ga}$ introduces additional levels in the band gap leading to ferromagnetic coupling due to the hybridization of the $p$ states of the Ga and N atoms with the Gd $d$ and $f$ states. Ferromagnetic exchange coupling energy of 76.4meV is obtained in presence of Gd-$V_{Ga}$ complex, and hence the FM is largely determined by the cation vacancy-rare earth complex defects in GaN NWs. On the other hand, the $V_N$, which introduce additional electron carriers, does not assist in increasing the ferromagnetic exchange energy.

[*] Corresponding author iman.roqan@kaust.edu.sa

I. Introduction

Gallium Nitride (GaN) nanowires (NWs) have several promising electronic and optoelectronic applications, ranging from light emitting diodes to sensing devices, high power transistors to spintronics, due to the wide and direct band gap and structural confinement properties [1-5]. Owing to the recent advancements in experimental techniques, it has been possible to successfully synthesize and characterize GaN NWs [1-3, 6]. Semiconductor NWs possess superior mechanical quality at high temperatures compared to bulk, rendering them more desirable candidates for spintronic devices [7]. It is well established that NW properties can be modified by changing the concentration of native defects as well as by the introduction of external impurities [8-11]. The most common dopants in GaN NWs include transition metals (TMs) and rare earth (RE) elements. The structural, magnetic and optical characteristics of TM [12, 13] and RE doped GaN NWs [14-16] are currently being investigated using a wide range of experimental techniques. The cylindrical geometry and large surface to volume ratio characterizing the NWs can enhance the spin polarization significantly, while reducing the formation energy ($E^f$) of defects [8, 17]. In addition, RE doped GaN has been the subject of intensive research focusing on bulk materials [18-24]. However, experimental behavior of the RE-doped GaN NWs is still not well understood and the theoretical studies that have been conducted thus far tend to focus solely on the effects of native defects and TM dopants [10, 11, 17, 25-27]. Theoretical studies can advance the understanding of the role of impurities and defects in explaining the experimental findings of GaN NWs. In this scenario, RE doped GaN NWs present an interesting research problem from the fundamental and technological point of view. In particular,

Gd is one of the promising RE dopants, reported to carry high magnetic moments and enhanced Curie temperatures in bulk GaN [18, 23, 24]. However, the mechanism of the magnetic and electronic behaviors of Gd doped GaN NW has not been investigated so far.

The diameters of NWs can be tuned from 1nm to as large as 200nms, the former possess distinguishing magnetic, optical and transport properties due to quantum confinement effects, compared to that of larger diameters [7, 9, 28]. Due to computational limitations, theoretical studies are often restricted to the modeling of NWs of few nm in diameter. However, recent advances in fabrication techniques have enabled the synthesize of NWs of small diameters (1.5 to 5 nm) [28, 29]; and therefore, density functional theory (DFT) studies have been successful in predicting comparable results [30]. Moreover, DFT studies on small-diameter nanowires assists in understanding the quantum confinement effects, which give rise to significant density of states (DOS) at the band edge [31]. In this work, we show that Gd doped GaN NWs have strong potential for practical applications in spintronic devices operating at room temperature. We report a systematic study, incorporating both native defects and Gd dopants in GaN NWs. GaN NW of 1 nm diameter is considered, which is adequate for investigating the relative stability and clustering effects of Gd in the dilute limit.

II. Methodology

The present study investigates a wurtzite $Ga_{48}N_{48}$ NW along the [0001] direction modeled using the supercell approach. The length of the supercell is $2c$, where $c$ is the bulk lattice parameter along the [0001] direction such that the simulation cell contains 4

GaN atomic layers with 96 atoms. In the planes normal to the growth direction, vacuum region of about 20 Å is used to ensure negligible interaction between the periodically repeated NW images. On the other hand, along the Z direction, infinite periodicity is maintained. The computations are performed using the *Vienna Abinitio Simulation Package* (VASP) [32, 33]. A cut off energy of 400 eV is used to expand the plane waves contained in the basis set. The pseudopotentials used are that of projected augmented wave (PAW) basis [34] with GGA in the Perdew Burke-Ernzerhof (PBE) form for exchange and correlations [34]. Atomic coordinates are relaxed within energy and force tolerances of 0.0001 eV and 0.004 eV/Å, respectively. The Brillouin zone is sampled using a 1×1×2 Monkhorst pack k-grid for the total energy calculations, whereas a larger k-grid of (1×1×8) is used for the electronic structure analysis. We employ the GGA+*U* method for our calculations, as it is well known that standard DFT may not accurately describe the electronic structure of GaN due to strong electronic correlations of the *3d* electrons. The parameters used in the calculations comprise on-site Coulomb parameter $U$ = 6.7 eV with a *J* value of 0.7 eV for Ga, and $U$ = 7.4 eV with a *J* value of 0.5 eV for Gd, in line with the Duradev's approach [35] as implemented in VASP.

III. Results and Discussion

Initially, the bulk unit cell parameters for wurtzite GaN is calculated for the purpose of comparison with experimental results. The lattice parameters *a* = 3.18 Å and *c* = 5.19 Å thus obtained agree well with the experimental values (*a* = 3.1876 Å; *c* = 5.1846 Å) [36], hence establishing the accuracy of our calculations. These optimized lattice parameters are used to construct the GaN NW and which is subjected to further optimization. It is important to obtain stable structure for the NW, due to the large surface area to volume ratio of the nanostructures. After the relaxation, the Ga-N bond length suffers distortion along the *c*-direction, and decreases to 1.85 Å compared to the bond length in bulk GaN (1.95 Å) [36]. The Ga-N bond length of the surface atoms are ~1.87 Å, in the *a-b* plane, while that of the inner atoms are ~ 1.92 Å. As expected, the surface atoms are characterized by shorter average bond lengths than the inner atoms, due to the reduced coordination numbers. In bulk GaN, the distance between the second nearest neighbor Ga atoms is 3.21 Å, while in the NW this is reduced to ~ 2.97 Å. The decrease in bond lengths obtained in NWs is a result of the increased surface to volume ratio, and similar trends have been reported in previous first principles studies as well [10, 37]. We compare the total densities of states (TDOS) of bulk GaN and pristine GaN NW calculated using the GGA+U approximation, and plotted in Figure 1. It is observed that the position of Fermi level ($E_F$) in bulk and in NW is almost identical, revealing the semiconducting characteristics of the NW, because $Gd^{3+}$ ions located at the Ga site is an isovalent impurity in GaN. The DOS of pristine NW shows zero spin-polarization, exhibiting a non-magnetic nature similar to the bulk GaN.

To analyze the possibility of dopant driven magnetism, one Gd atom is incorporated in to Ga sites in the GaN NW. This is in accordance with experimental findings that RE dopants preferentially occupy Ga sites in GaN [38]. The calculations are performed for 12 non-equivalent configurations, grouped into three categories to encompass the bulk-like, subsurface and surface positions, allowing us to investigate the more stable location of Gd atoms in NWs, as shown in Figure 3. Relaxation without spin or geometrical constraints is carried out to identify the most stable configuration. The $E^f$ of these NW configurations are calculated according to,

$$E^f = (E^d(total) - E(Ga_{48}N_{48}) + n_i\mu_i(Ga) - n_i\mu_i(Gd))/n_i \quad \ldots\ldots\ldots \quad (1);$$

where $E^d$ (*total*) and $E$ ($Ga_{48}N_{48}$) denote the total energy of the NW containing Gd and total energy of the pristine GaN NW, respectively; $n_i$ represents number of atoms removed/added for respective elements; $\mu_i$ is the chemical potential and which is the total energy calculated for metallic Ga and Gd elements.

The $E^f$ for each configuration is presented in Table 1, which shows that the most stable Gd configuration is obtained at the surface of the NW. This implies the effect of self-purification observed in nanomaterials [39]. When the impurities and defects require high energy to be incorporated in to the nanostructure, they are expelled to the surface. In this case, the impurity stability is low inside the NW due to the fact that impurity needs greater energy to occupy deep states in the NW. This phenomenon is also observed in experimental findings pertaining to the migration of impurities to the surface of the NWs [40].

We further analyze the magnetic properties for undoped and Gd-doped GaN NWs. When Gd impurities (which carry a magnetic moment of 6.9 $\mu_B$/atom) are doped in to GaN NW, replacing the Ga site, no magnetic moment is induced to the nearby Ga or N sites. Moreover, the total moment is equivalent to that of Gd atoms due to the strong localization of the partially filled Gd *5d* and *4f* sublevels. The electronic structure of Gd doped GaN NW is calculated and shown in Figure 5 (top panel). The $E_F$ is situated near valence band maximum (VBM), indicating that the $Ga_{47}N_{48}Gd$ NW exhibit

*p*-type behavior. The relatively weak ferromagnetic exchange interaction between the Gd atoms can be explained by the electronic structure characteristics. More specifically, the majority 4*f* orbitals of Gd atoms are located well below VBM, and the minority 4*f* orbitals are situated above the conduction band minimum (CBM). The majority 4*f* levels are completely filled, while the minority levels are completely empty. The Gd 5*d* orbitals are resonant with the 4*f* states and appear at the same energy intervals. The magnetic moment of 6.9$\mu_B$ for the Gd-doped GaN NW originates from these 4*f* orbitals, which results in the reduction in the magnetic moment of Gd compared to that of metallic Gd. In addition, the localized 4*f* electrons in the VB are involved in the chemical bonding, ensuring that the system remains semiconducting. Therefore, we conclude that Gd-doped defect-free GaN NWs are paramagnetic. This assertion is in accordance with the findings reported by Ney *et al* [21], wherein the paramagnetic nature of Gd doped GaN is inferred from magnetic measurements, and the presence of additional carriers are proposed to stabilize RTFM.

Since the Gd atom alone is not sufficient to establish RTFM in the NWs, the influence of Gd atoms along with native point defects such as Ga and N vacancy ($V_{Ga}$ and $V_N$ respectively) in the vicinity is analyzed. Prior to this, we have introduced $V_{Ga}$ and $V_N$ in pristine GaN NW, and calculated the $E^f$ to examine their stability. The $E^f$ is calculated using the following expression:

$E^f = (E^d(\text{total}) - E(\text{Ga}_{48}\text{N}_{48}) + n_i\mu_i )/n_i$    ……….(2);

where $E^d(\text{total})$ and $E(\text{Ga}_{48}\text{N}_{48})$ are the total energy of the supercell containing the defect and that of a perfect supercell, respectively; whereas $n_i$ and $\mu_i$ denote the number of atoms removed and chemical potential in different thermodynamic limits for respective elements.

The calculated $E^f$ for $V_{Ga}$ and $V_N$ are 5.77 eV and 6.42 eV respectively (see Table 2), which are lower than the values calculated for bulk GaN. The $V_{Ga}$ is favorably stabilized in GaN NW relative to $V_N$, contrary to bulk GaN, where $V_N$ forms more easily than $V_{Ga}$ [41]. This finding shows that it is easier to form localized defect bands states in GaN NW, compared to bulk GaN contributed by $V_{Ga}$. The lower defect formation energies are due to the enhanced degree of freedom for relaxation surrounding the vacancies in the NW compared to bulk, which in turn gives rise to a decrease in the defect-induced local strain in the vicinity. Following the structural relaxation, defect states become more localized, enhancing the spin polarization of the NW.

Incorporating one $V_{Ga}$ in pristine NW results in a total magnetic moment of 2.0$\mu_B$ and the N atoms near the vacancy site acquire magnetic moments of 0.44 to 0.68 $\mu_B$/atom, owing to the bonding characteristics of the wurtzite structure. However, $V_N$ induces negligible magnetic polarization (~ 0.005 $\mu_B$/atom) in the GaN NW. For further analysis, the TDOS calculated for GaN NW with $V_{Ga}$ and $V_N$ is shown in Figure 2. Creating $V_G$ introduces three holes in effect, and it is observed that $V_{Ga}$ pushes the VBM towards $E_F$. Two unoccupied minority spin states are created in presence of $V_{Ga}$ and the splitting of majority and minority DOS occurs, which gives rise to magnetic polarization in the NW. Since $V_{Ga}$ introduces additional levels in the band gap, and being more stable defect in the NW, can invoke magnetism mediated by holes. Whereas, $V_N$ is least stable and do not introduce any additional levels in the band gap. The electron carriers introduced by creating $V_N$ result in shifting $E_F$ below CBM, introducing a deep state which is situated well below the CBM.

Inspired by the low $E^f$ of $V_{Ga}$ and favorable magnetism induced by it, we have extended the calculations by creating $V_{Ga}$ in the Gd doped NW. For comparison, calculations are also carried out for $V_N$ in the Gd doped NW and $E^f$ of the defects are calculated in each case. The $E^f$ results show that Gd atoms stabilize the vacancies ($V_{Ga}$ and $V_N$) more than in pristine NW, as can be seen from Table 2. This indicates that Gd assists in the stabilization of vacancies in GaN NW. As the ion radii of $Gd^{3+}$ are considerably greater than that of the host $Ga^{3+}$ ion, a high density of vacancies can be achieved in the NWs.

When one $V_{Ga}$ is introduced to the Gd doped NW, nearby N atoms get polarized. In addition, the maximum local magnetic moments at the N sites are increased from 0.68 to 0.88 $\mu_B$, whereas the total magnetic moment of the nanowire supercell is 9 $\mu_B$. The Gd moment also increases slightly to 7.14 $\mu_B$. The isosurface plot ($\rho = 9\times10^{-4}$ e/a.u.$^3$) for this configuration is presented (Figure 4(b)), which shows that the spin density is not only confined to the nearest neighbors, but also extends to the next nearest neighbors. The spin density calculated for NW containing $V_{Ga}$ in the absence of Gd is also presented for comparison (Figure 4(a)). It can be seen that the spin polarization extends to a greater number of N sites when both Gd atoms and $V_{Ga}$ exist in the vicinity. This indicates the increased hybridization between Gd and the defect bands. Therefore, magnetization can be sustained and stabilized in GaN NW in presence of sufficient number of $V_{Ga}$-Gd complexes that can polarize the GaN matrix efficiently. The defect wave functions due to the $V_{Ga}$ extends through the N sites, owing to the bonding arrangement of wurtzite GaN, enhancing the ferromagnetic coupling in GaN NWs [19]. Hence, with the introduction of one $V_{Ga}$ with Gd impurity, the total magnetic moment undergoes an increase of about 2 $\mu_B$. This indicates that the occurrence of large magnetic moments in Gd doped GaN is due to the presence of cation vacancy- Gd dopant complexes.

To understand the mechanism behind this magnetism, we have calculated the electronic structure of Gd doped GaN NWs in presence of both $V_{Ga}$ and $V_N$, as shown in Figure 5(middle and bottom panels respectively). The DOS for the Gd doped GaN NW without any vacancy is plotted in Figure 5 (top panel). It can be noted that $V_{Ga}$ introduces a defect state in the minority spin near the $E_F$ and below the VBM, whereas positions of 4*f* states remain unaltered. It is well known that hybridization of defect levels with that of impurity atom near the $E_F$ and below the VBM leads to ferromagnetic coupling in Gd-doped GaN. The inset in the middle panel of Figure 5 shows the projected density of states of the hybridization of the defect band introduced by $V_{Ga}$ with Gd *4f* and *5d* states for the $Ga_{46}N_{48}Gd$ NW. It can be seen that the additional states near the $E_F$ are contributed by the hybridization between the Ga *d* and N *p* states and Gd (*d* and *f*) states in the spin down channel. The Gd *d* and *f* states in the VB show a clear spin up - spin down splitting. In the presence of $V_{Ga}$, spins introduced by these additional defect levels interact with the Gd spin ferromagnetically, resulting in a slight Gd magnetic moment increase, simultaneously polarizing the nearby N atoms, thus increasing the total magnetic moment of the system by about 3 $\mu_B$. Hence, possibility of a *p-d* exchange interaction arises between *d* orbitals of Ga and Gd and *p* orbitals of N. However, the presence of delocalized carriers is less likely to induce magnetism and hence zener's direct exchange mechanism [42] cannot be the cause of the observed FM in GaN NWs. In the present context, the magnetic polarization caused by the $V_{Ga}$ at the N sites helps in stabilizing FM in the presence of Gd. However, it is not possible for the *s-f* exchange interactions to occur, as the Gd atoms stay apart and do not prefer to aggregate. The Gd *f* and *d* states overlap well with defect band near the $E_F$ and VBM as shown in the inset in the middle panel of Figure 5. Therefore, our study establishes the idea of cation vacancy induced FM in RE doped GaN NWs due to the increased hybridization between the *p* states of Ga and N with *d* and *f* states of Gd [22]. Hence, the mechanism behind FM in Gd doped GaN NWs can be ferromagnetic superexchange-like interactions [19] mediated through the N sites.

To examine the strength of ferromagnetic exchange coupling and to demonstrate the existence of long-range magnetic interactions, we have introduced two Gd atoms, removing two Ga atoms from the NW supercell. Introducing more than one Gd atom will also give favorable insight into the clustering tendency of dopants in the NW. The $E^f$ for a pair of Gd atoms is calculated according to Eq. 1 and the results are plotted in Figure 6. Interestingly, the $E^f$ results show that no Gd aggregation can take place in the NWs. The Gd atoms prefer to stay apart as the $E^f$ decreases with the increase in distance (D) between Gd atoms, and shown in Figure 6. This observation is in line with experimental findings indicating that Gd atoms distribute homogeneously in wurtzite GaN without clustering [4, 18]. Further, to investigate the stability of RTFM, the total energy difference (ΔE) between configurations in which two Gd atoms are arranged in the ferromagnetic (FM) and antiferromagnetic (AFM) configurations (ΔE = $E_{FM}$ − $E_{AFM}$) are calculated. A positive value of ΔE implies that Gd atoms prefer FM coupling in the NW and vice versa. The calculated value of ΔE is only about 1.1 meV, signifying the paramagnetic nature of Gd-Gd interactions in the $Ga_{46}N_{48}Gd_2$ NW. However, introducing one hole (additional positive charge) by reducing the total number of valence electrons results in weak FM coupling, with ΔE of 12.4 meV, which agrees with findings in bulk GaN, where the presence of holes introduced by the $V_{Ga}$ is shown to enhance the ferromagnetic coupling [19].

Further, when one $V_{Ga}$ is introduced into the $Ga_{46}N_{48}Gd_2$ NW, a significant enhancement in magnetic coupling is observed, as one $V_{Ga}$ introduces three holes. These holes introduced by the cation vacancy are in a triplet magnetic ground state and enhance the FM of the system. The calculated value of ΔE is 76.4 meV, which is strong enough to establish RTFM coupling, since a minimum ΔE of 30 meV is required to establish magnetic coupling against thermal fluctuations. We further observe that strength of ferromagnetic coupling increases with the presence of $V_{Ga}$ in the nearby atomic sites, as has been reported for bulk GaN [19]. On the other hand, negligible magnetization is obtained with the introduction of $V_N$ and electron carriers, in accordance with previous studies, which showed that $V_N$ and electron carriers are not involved in mediating ferromagnetic interactions in Gd:GaN [22]. The behavior of $V_N$ in Gd doped GaN can be explained from the TDOS as shown in Figure 5 (bottom panel). In presence of $V_N$, the VBM shifts away from the $E_F$ and the GaN NW shows semi-insulating nature. The DOS shows that the 4*f* majority levels are shifted well below the VBM, whereas the minority spin states move near the $E_F$. However, at $E_F$, spin up and spin down DOS are symmetric, giving rise to zero spin polarization in the system. Thus, the proposition that electron mediated FM in Gd-GaN stems from *s-f* or *s-d* coupling interaction [23] is not valid in the present context, as the possible origin of electron carriers is $V_N$, and such vacancies are not favorably forming in the NW and induce zero spin polarization. Hence it can be concluded that the large magnetic moment contributed by the Gd atom combined with the spin polarization induced by the $V_{Ga}$ to the N sites strengthens the magnetic coupling in the NW.

IV. Conclusions

To summarize, investigations on GaN NWs in presence of intrinsic and extrinsic defects are carried out. Gd atoms do not prefer to cluster in the NW and exhibit paramagnetic coupling in the absence of carriers. Point defect formation energy is reduced in the GaN NW compared to bulk. Moreover, $V_{Ga}$ is more stable than $V_N$ and induces FM in the NW. The Gd atoms reduce the $E^f$ of point defects and FM is strengthened by the Gd-$V_{Ga}$ complex defects. Our study thus emphasizes the importance of the presence of additional hole carriers in establishing RTFM in Gd doped GaN NWs.

# Reference

[1] Li P G, Guo X, Wang X and Tang W H 2009 Single-crystalline wurtzite GaN nanowires and zigzagged nanostructures fabricated by sublimation sandwich method *Journal of Alloys and Compounds* **475** 463-8

[2] Kim J-R, So H M, Park J W, Kim J-J, Kim J, Lee C J and Lyu S C 2002 Electrical transport properties of individual gallium nitride nanowires synthesized by chemical-vapor-deposition *Applied Physics Letters* **80** 3548-50

[3] Chen C-Y, Zhu G, Hu Y, Yu J-W, Song J, Cheng K-Y, Peng L-H, Chou L-J and Wang Z L 2012 Gallium Nitride Nanowire Based Nanogenerators and Light-Emitting Diodes *ACS Nano* **6** 5687-92

[4] Asahi H, Zhou Y K, Hashimoto M, Kim M S, Li X J, Emura S and Hasegawa S 2004 GaN-based magnetic semiconductors for nanospintronics *Journal of Physics: Condensed Matter* **16** S5555

[5] Westover T, Jones R, Huang J Y, Wang G, Lai E and Talin A A 2009 Photoluminescence, Thermal Transport, and Breakdown in Joule-Heated GaN Nanowires *Nano Letters* **9** 257-63

[6] Li J and He M 2014 One-Step In Situ Direct Growth of GaN Nanowire Devices *Science of Advanced Materials* **6** 699-702

[7] Yan R, Gargas D and Yang P 2009 Nanowire photonics *Nat Photon* **3** 569-76

[8] Aravindh S A, Schwingenschloegl U and Roqan I S 2014 Ferromagnetism in Gd doped ZnO nanowires: A first principles study *Journal of Applied Physics* **116** 233906

[9] Assa Aravindh S, Mathi Jaya S, Valsakumar M C and Sundar C S 2012 Compositional variation of magnetic moment, magnetic anisotropy energy and coercivity in Fe(1−x)M x (M = Co/Ni) nanowires: an ab initio study *Appl Nanosci* **2** 409-15

[10] Carter D J, Gale J D, Delley B and Stampfl C 2008 Geometry and diameter dependence of the electronic and physical properties of GaN nanowires from first principles *Physical Review B* **77** 115349

[11] Carter D J and Stampfl C 2009 Atomic and electronic structure of single and multiple vacancies in GaN nanowires from first-principles *Physical Review B* **79** 195302

[12] Stamplecoskie K G, Ju L, Farvid S S and Radovanovic P V 2008 General Control of Transition-Metal-Doped GaN Nanowire Growth: Toward Understanding the Mechanism of Dopant Incorporation *Nano Letters* **8** 2674-81

[13] Radovanovic P V, Barrelet C J, Gradečak S, Qian F and Lieber C M 2005 General Synthesis of Manganese-Doped II−VI and III−V Semiconductor Nanowires *Nano Letters* **5** 1407-11

[14] Cao Y P, Shi F, Xiu X W, Sun H B, Guo Y F, Liu W J and Xue C 2010 Syntheses and properties of Tb-doped GaN nanowires *Inorg Mater* **46** 1096-99

[15] Cao Y P, Shi F, Sun H B, Liu W J, Guo Y F and Xue C S 2010 Growth and properties of Dy-doped GaN nanowires *The European Physical Journal - Applied Physics* **50**

[16] Lorenz K, Nogales E, Miranda S M C, Franco N, Méndez B, Alves E, Tourbot G and Daudin B 2013 Enhanced red emission from praseodymium-doped GaN nanowires by defect engineering *Acta Materialia* **61** 3278-84

[17] Dev P, Zeng H and Zhang P 2010 Defect-induced magnetism in nitride and oxide nanowires: Surface effects and quantum confinement *Physical Review B* **82** 165319

[18] Dhar S, Brandt O, Ramsteiner M, Sapega V F and Ploog K H 2005 Colossal Magnetic Moment of Gd in GaN *Physical Review Letters* **94** 037205

[19] Gohda Y and Oshiyama A 2008 Intrinsic ferromagnetism due to cation vacancies in Gd-doped GaN: First-principles calculations *Physical Review B* **78** 161201

[20] Mitra C and Lambrecht W R L 2009 Interstitial-nitrogen- and oxygen-induced magnetism in Gd-doped GaN *Physical Review B* **80** 081202

[21] Ney A, Kammermeier T, Ney V, Ye S, Ollefs K, Manuel E, Dhar S, Ploog K H, Arenholz E, Wilhelm F and Rogalev A 2008 Element specific magnetic properties of Gd-doped GaN: Very small polarization of Ga and paramagnetism of Gd *Physical Review B* **77** 233308

[22] Liu L, Yu P Y, Ma Z and Mao S S 2008 Ferromagnetism in GaN:Gd: A Density Functional Theory Study *Physical Review Letters* **100** 127203

[23] Dalpian G M and Wei S-H 2005 Electron-induced stabilization of ferromagnetism in GaGdN *Physical Review B* **72** 11520

[24]	Dev P, Xue Y and Zhang P 2008 Defect-Induced Intrinsic Magnetism in Wide-Gap III Nitrides *Physical Review Letters* **100** 117204
[25]	Wang Q, Sun Q, Jena P and Kawazoe Y 2005 Ferromagnetic GaN−Cr Nanowires *Nano Letters* **5** 1587-90
[26]	Tsai M-H, Jhang Z-F, Jiang J-Y, Tang Y-H and Tu L W 2006 Electrostatic and structural properties of GaN nanorods/nanowires from first principles *Applied Physics Letters* **89** 203101
[27]	Gulans A and Tale I 2007 Ab initio calculation of wurtzite-type GaN nanowires *physica status solidi (c)* **4** 1197-200
[28]	Schio P, Vidal F, Zheng Y, Milano J, Fonda E, Demaille D, Vodungbo B, Varalda J, de Oliveira A J A and Etgens V H 2010 Magnetic response of cobalt nanowires with diameter below 5 nm *Physical Review B* **82** 094436
[29]	Shen G, Liang B, Wang X, Huang H, Chen D and Wang Z L 2011 Ultrathin $In_2O_3$ Nanowires with Diameters below 4 nm: Synthesis, Reversible Wettability Switching Behavior, and Transparent Thin-Film Transistor Applications *ACS Nano* **5** 6148-55
[30]	Xu Z, Zheng Q-R and Su G 2012 Charged states and band-gap narrowing in codoped ZnO nanowires for enhanced photoelectrochemical responses: Density functional first-principles calculations *Physical Review B* **85** 075402
[31]	Sheetz R M, Ponomareva I, Richter E, Andriotis A N and Menon M 2009 Defect-induced optical absorption in the visible range in ZnO nanowires *Physical Review B* **80** 195314
[32]	Kresse G and Furthmüller J 1996 Efficient iterative schemes for ab initio total-energy calculations using a plane-wave basis set *Physical Review B* **54** 11169-86
[33]	Kresse G and Hafner J 1993 Ab initio molecular dynamics for liquid metals *Physical Review B* **47** 558-61
[34]	Blöchl P E 1994 Projector augmented-wave method *Physical Review B* **50** 17953-79
[35]	Dudarev S L, Botton G A, Savrasov S Y, Humphreys C J and Sutton A P 1998 Electron-energy-loss spectra and the structural stability of nickel oxide: An LSDA+U study *Physical Review B* **57** 1505-9
[36]	Porowski S 1998 Bulk and homoepitaxial GaN-growth and characterisation *Journal of Crystal Growth* **189–190** 153-8
[37]	Wang Z, Li J, Gao F and Weber W J 2010 Defects in gallium nitride nanowires: First principles calculations *Journal of Applied Physics* **108** 044305
[38]	Wahl U, Alves E, Lorenz K, Correia J G, Monteiro T, De Vries B, Vantomme A and Vianden R 2003 Lattice location and optical activation of rare earth implanted GaN *Materials Science and Engineering: B* **105** 132-40
[39]	Dalpian G M and Chelikowsky J R 2006 Self-Purification in Semiconductor Nanocrystals *Physical Review Letters* **96** 226802
[40]	Ungureanu M, Schmidt H, Xu Q, von Wenckstern H, Spemann D, Hochmuth H, Lorenz M and Grundmann M 2007 Electrical and magnetic properties of RE-doped ZnO thin films (RE = Gd, Nd) *Superlattices and Microstructures* **42** 231-5
[41]	Limpijumnong S and Van de Walle C G 2004 Diffusivity of native defects in GaN *Physical Review B* **69** 035207
[42]	Dietl T, Ohno H, Matsukura F, Cibert J and Ferrand D 2000 Zener Model Description of Ferromagnetism in Zinc-Blende Magnetic Semiconductors *Science* **287** 1019-22

Figure Captions

Figure 1. The total DOS corresponding to bulk GaN and pristine $Ga_{48}N_{48}$ NW.

Figure 2. The TDOS of GaN NW with one $V_{Ga}$ ($Ga_{47}N_{48}$) and with one $V_N$ ($Ga_{48}N_{47}$).

Figure 3. The GaN NW with Gd doping. The twelve non-equivalent doping sites are grouped into three and each location is numbered as indicated in the figure. The Ga and N atoms are indicated by pink and dark yellow colors, respectively. The three non-equivalent Gd sites are represented by bright yellow, blue and silver colors, respectively.

Figure 4. The spin density plot for (a) GaN NW in presence of $V_{Ga}$ ($Ga_{47}N_{48}$) and (b) Gd-doped GaN NW in presence of $V_{Ga}$ ($Ga_{46}N_{48}Gd$). The green and white spheres represent Ga and N atoms, respectively, whereas the red spheres in (b) represents Gd atom.

Figure 5. The TDOS of Gd doped GaN NW ($Ga_{47}N_{48}Gd$), Gd doped NW with $V_{Ga}$ ($Ga_{46}N_{48}Gd$) and with $V_N$ ($Ga_{47}N_{47}Gd$), along with projected DOS of Gd $d$ and $f$ orbitals. The inset in the middle panel shows the projected DOS of Ga and Gd atoms in $Ga_{46}N_{48}Gd$ NW.

Figure 6. The formation energy (in eV) for two Gd atoms in GaN NW ($Ga_{46}N_{48}Gd_2$) with their distance of separation (in Angstrom).

Table 1. The Formation energy ($E^f$) for the 12 NW configurations presented in Figure 3

| Configuration | $E^f$ (eV) |
|---|---|
| G1-a | -2.477 |
| G1-b | -2.480 |
| G1-c | -2.477 |
| G1-d | -2.475 |
| G2-a | -1.367 |
| G2-b | -1.366 |
| G2-c | -1.366 |
| G2-d | -1.365 |
| G3-a | -1.372 |
| G3-b | -1.367 |
| G3-c | -1.373 |
| G3-d | -1.372 |

Table 2. The Formation energy ($E^f$) of $V_{Ga}$ and $V_N$ with and without Gd dopant in the GaN NW.

| Configuration | $E^f$ (eV) |
|---|---|
| Ga$_{47}$N$_{48}$ | 5.77 |
| Ga$_{46}$N$_{48}$Gd | 4.82 |
| Ga$_{48}$N$_{47}$ | 6.42 |
| Ga$_{47}$N$_{47}$Gd | 5.41 |

Figure 1.

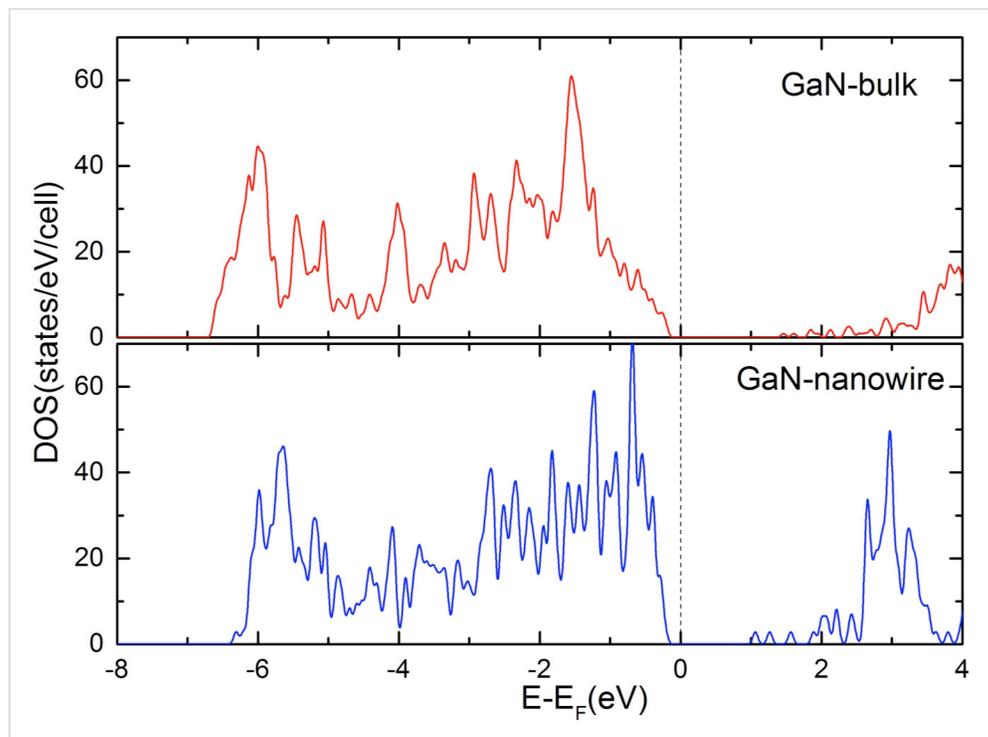

Figure 2.

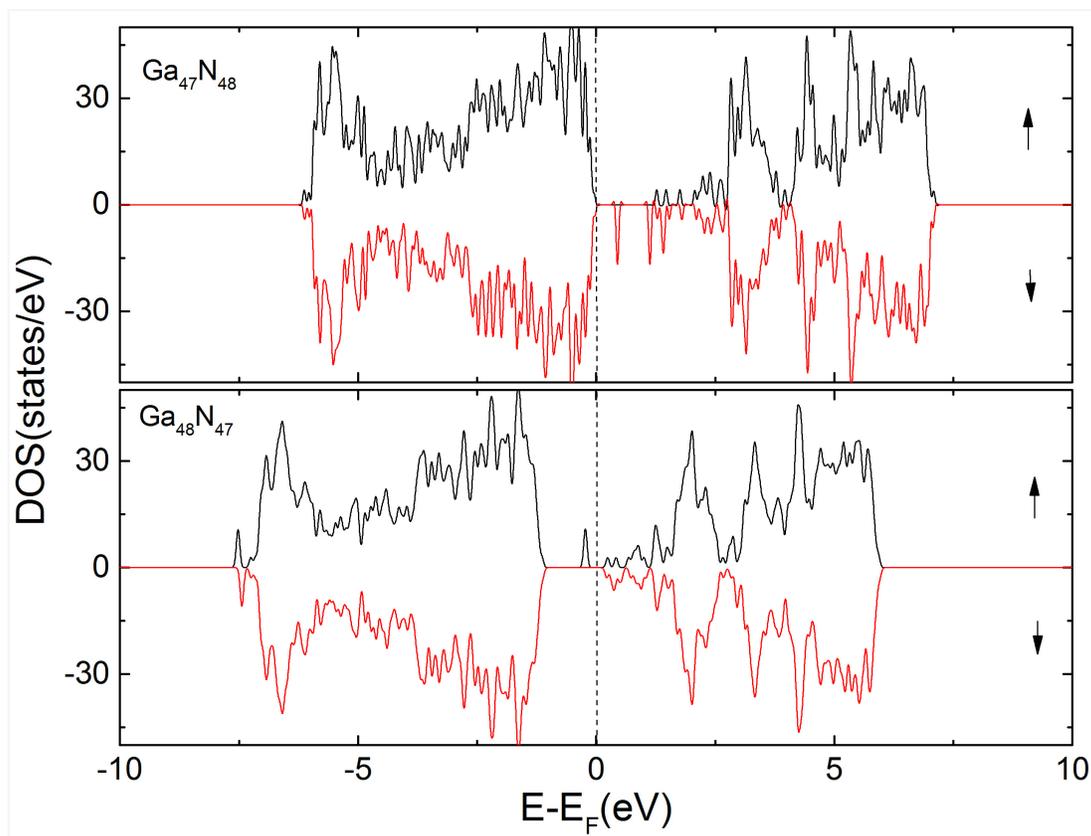

Figure 3.

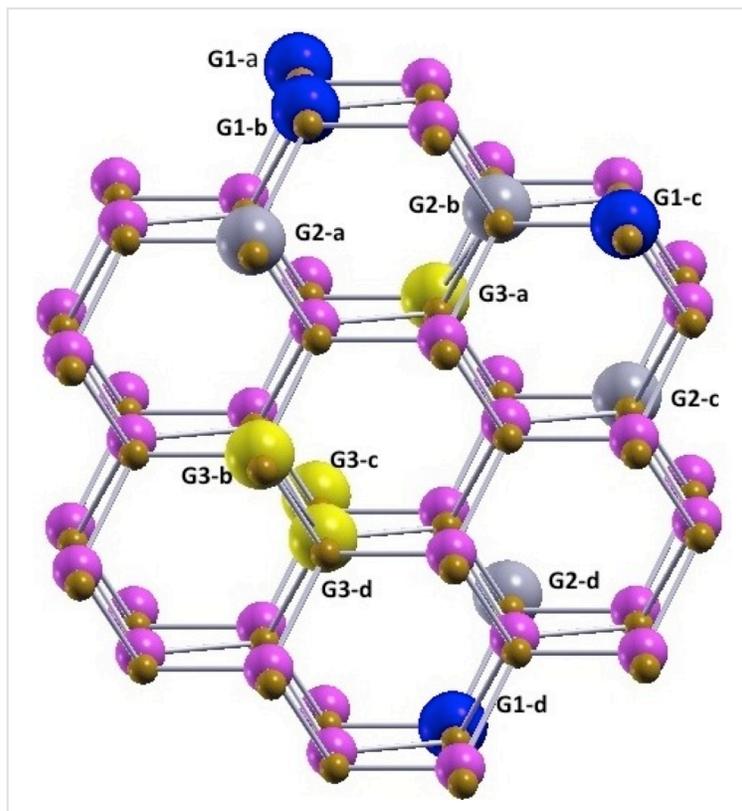

Figure 4.

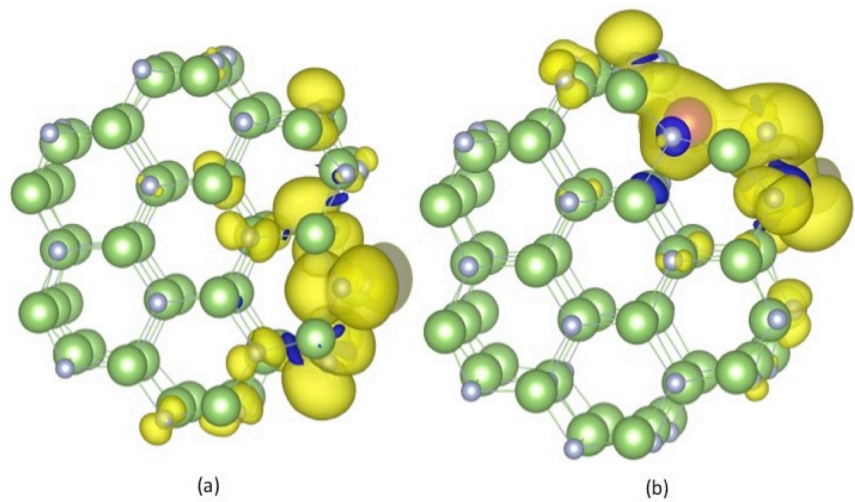

Figure 5.

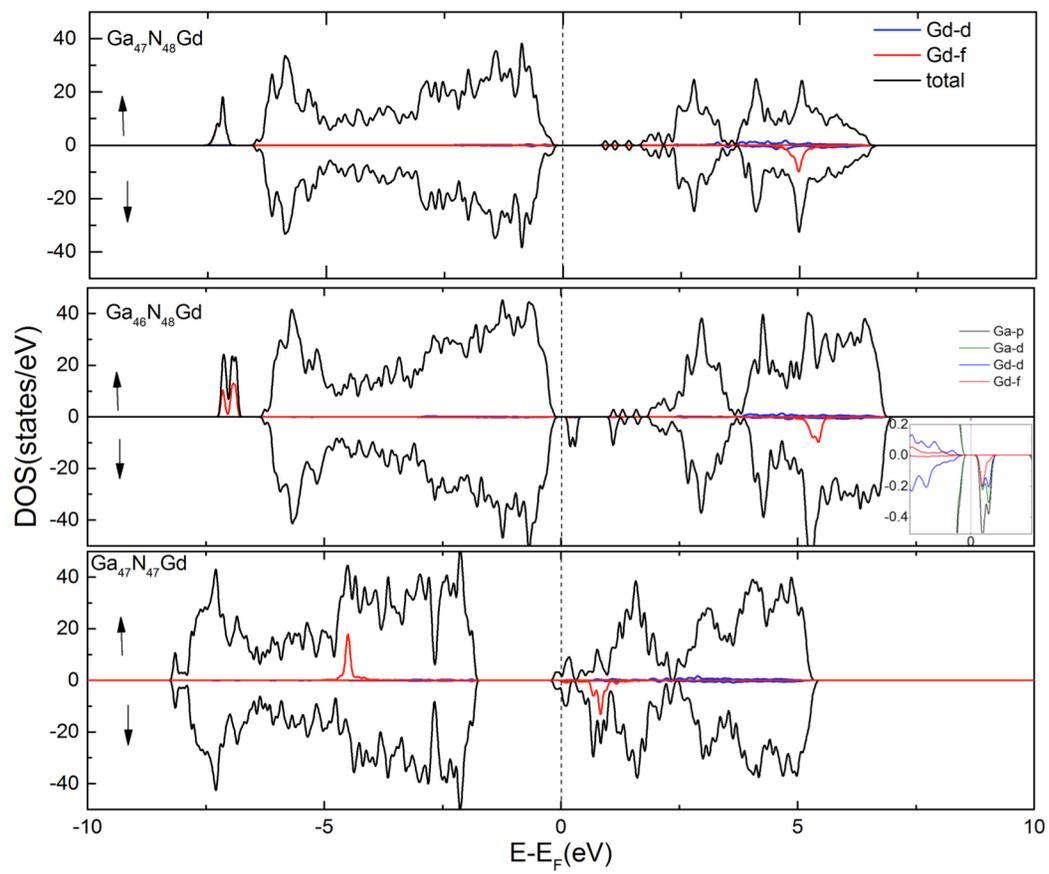

Figure 6.

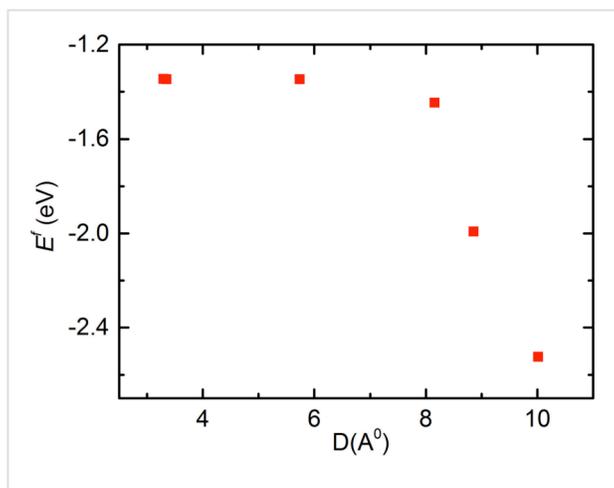